\documentclass[aip,pop,numerical,preprint,endfloats]{revtex4-1}
\usepackage{CJK}
\usepackage{graphicx}
\usepackage{dcolumn}
\usepackage{bm}
\usepackage{amsmath}
\usepackage{amssymb}

\begin{document}
\begin{CJK*}{GB}{gbsn}

\title{NIMROD calculations of energetic particle driven toroidal Alfv\'en eigenmodes}


\author{Yawei Hou (ºîÑÅΡ)}
\affiliation{CAS Key Laboratory of Geospace Environment and Department of Modern Physics, University of Science and Technology of China, Hefei, Anhui 230026, China}

\author{Ping Zhu (Öìƽ)}\email{pzhu@ustc.edu.cn}
\affiliation{CAS Key Laboratory of Geospace Environment and Department of Modern Physics, University of Science and Technology of China, Hefei, Anhui 230026, China}
\affiliation{KTX Laboratory and Department of Modern Physics, University of Science and Technology of China, Hefei, Anhui 230026, China}
\affiliation{Department of Engineering Physics, University of Wisconsin-Madison, Madison, Wisconsin 53706, USA}

\author{Charlson C. Kim (½ðÖÇÉÆ)}
\affiliation{SLS2 Consulting, San Diego, California 92107, USA}

\author{Zhaoqing Hu (ºú³¯Çå)}
\affiliation{Dalian University of Technology, Dalian, Liaoning 116024, China}

\author{Zhihui Zou (×ÞÖ¾»Û)}
\affiliation{CAS Key Laboratory of Geospace Environment and Department of Modern Physics, University of Science and Technology of China, Hefei, Anhui 230026, China}

\author{Zhengxiong Wang (ÍõÕýÐÚ)}
\affiliation{Dalian University of Technology, Dalian, Liaoning 116024, China}

\author{the NIMROD Team}


\date{\today}

\begin{abstract}
Toroidal Alfv\'en eigenmodes (TAEs) are gap modes induced by the toroidicity of tokamak plasmas in absence of continuum damping. They can be excited by energetic particles (EPs) when the EP drive exceeds other dampings. A TAE benchmark case, which was proposed by the International Tokamak Physics Activity (ITPA) group, is studied in this work. Numerical calculations of linear growth of TAEs driven by EPs in a circular-shaped, large aspect ratio tokamak have been performed using the Hybrid Kinetic-MHD (HK-MHD) model implemented in the NIMROD code. This HK-MHD model couples a $\delta f$ particle-in-cell (PIC) representation of EPs with the 3D MHD representation of the bulk plasma through moment closure for the momentum conservation equation. Both the excitation of TAEs and their transition to energetic particle modes (EPMs) have been observed. The influence of EP density, temperature, density gradient and position of the maximum relative density gradient, on the frequency and the growth rate of TAEs are obtained, which are consistent with those from eigen-analysis calculations and gyrokinetic simulations for an initial Maxwellian distribution of EPs. The relative pressure gradient of EP at the radial location of TAE gap, which represents the drive strength of EPs, can strongly affect the growth rate of TAEs. When the density and temperature of EP distribution are above certain threshold, the transition from TAE to EPM occurs and the mode structure changes.
\end{abstract}


\maketitle
\end{CJK*}

\section{Introduction}
Eigenmodes of Alfv\'en Waves can be excited by energetic particles (EPs) generated from neutral beam injection or other heating processes in tokamaks\cite{Rosenbluth1975,Wong1999,Sharapov2013,Chen2016}, as well as fusion produced alpha particles. In burning plasma devices\cite{Zonca2006} like ITER\cite{Fasoli2007}, the drive of EPs will become very strong because the velocity of alpha particle is close to be Alfv\'enic. In regions without strong continuum damping\cite{Berk1991,Rosenbluth1992}, Toroidal Alfv\'en Eigenmodes (TAEs) can be easily excited\cite{Berk1993}. Meanwhile, the destabilized TAEs can induce the transport of EPs and degrade the confinement of burning plasmas. The interaction between TAEs and EPs in tokamak plasmas, especially the burning plasmas, has been a topic of primary interest in fusion research.

Linear theory of TAEs has been developed for various circumstances\cite{Chen2016}, including large\cite{Chen1994,Fu1989a} and small\cite{Fu1989b} aspect ratios, ideal\cite{Rosenbluth1992} and nonideal\cite{Candy1993,Candy1994} MHD models, low-$n$\cite{Cheng1986,Gorelenkov1999} and hign-$n$ modes\cite{Rosenbluth1992}, etc. Here, $n$ is the toroidal mode number. Cheng et al \cite{Cheng1986} solve the linear ideal MHD equations to study the low-$n$ shear Alfv\'en spectra in axisymmetric plasmas and find that toroidicity can break up the continuous spectra and form an continuum gap, within which an AE can be excited. Fu et al \cite{Fu1989a} consider kinetic effects of alpha particle and electron on TAEs in circular flux surface tokamak using the linear drift kinetic equation coupled with ideal MHD equations. They find that the low-$n$ TAEs can be destablized by alpha particles. Rosenbluth et al \cite{Rosenbluth1975} point out that the shear Alfv\'en wave can be destablized by energetic ion and develop asymptotic formulas\cite{Rosenbluth1992} to calculate the continuum damping rate for high-$n$ TAE in a low-bata, large-aspect-ratio, circular tokamak. Chen \cite{Chen1994} studied the high-$n$ TAE and KBM (Kinetic Ballooning Mode) using the ballooning mode representation of the eigenmode equation in large-aspect-ratio circular tokamak. With the nonperturbative EPs, both gap modes (TAE and KBM) and EP continuum mode (EPM) are found from the dispersion relation. Berk et al \cite{Berk1993} and Candy et al \cite{Candy1993, Candy1994} extend the TAE study to nonideal MHD regimes (finite ion Larmor radius, electron inertia, and collisions) for arbitrary mode numbers in large-aspect-ratio circular tokamaks.

The linear properties of shear Alfv\'en waves (SAWs), such as the excitation of TAEs, are relatively well understood\cite{Chen2016} and the methods for identifying SAWs in experiment\cite{Fasoli2000} are well developed. The competition between the energetic particle drive and different dampings\cite{Berk1992,Zonca1992,Fu2005,Lauber2005}, as well as the redistribution of EPs in phase space and the transport of EPs in real space, can cause complex behaviors of Alfv\'en eigenmodes\cite{Appel1995}. More efforts are to be devoted to numerical simulations in order to understand the nonlinear effects of SAWs on bulk plasmas and the transport of energetic particles, especially the large-amplitude collective instabilities leading to energetic particle redistributions in real and phase spaces\cite{Fasoli2007}.

To investigate the global and nonlinear features of TAEs driven by EPs, various numerical calculations and simulations have been performed\cite{Fu2005, Lauber2007, Todo2012, Briguglio2007, Nishimura2007, Mishchenko2008, Nishimura2009}.
Fu\cite{Fu2005} investigates the kinetic damping of TAE in JET using reduced kinetic equations and finds the continuum damping important in the edge plasma. LIGKA is a linear gyrokinetic code developed by Lauber et al\cite{Lauber2007} to study the EP effect on MHD stability, as well as the drive and damping of TAEs.
Todo et al\cite{Todo2012} develop and use MEGA to study TAE bursts in the plasma regimes of the TFTR tokamak. Using HMGC code, Briguglio et al\cite{Briguglio2007} study the large-amplitude, bursting Alfv\'en modes in JT-60U, especially the dynamics of EP in phase space. In addition, global  gyrokinetic PIC codes, GTC and GYGLES, have been applied to the study of TAEs\cite{Nishimura2007,Mishchenko2008}. Besides those simulations in large aspect ratio and low $\beta$ tokamak plasmas using reduced MHD models, other codes, such as M3D~\cite{Park1999} and NIMROD~\cite{Sovinec2004}, have been developed to study EP physics in realistic tokamak configurations based on full MHD models\cite{Kim2004,Lang2010}.

In this paper, we report on the kinetic MHD simulations of SAWs driven by fast
particles in a tokamak using the hybrid kinetic-MHD (HK-MHD) model in the NIMROD code\cite{Kim2004}. The kinetic effects of fast minority ions are included in the HK-MHD model through the addition of fast particle pressure to the momentum conservation equation, which is computed using a $\delta f$ particle-in-cell (PIC) algorithm on a quadrilateral finite element grid\cite{Kim2004,Kim2008,Brennan2012,Takahashi2009}. In preparation for studying the nonlinear EP driven SAWs in realistic tokamak plasmas, we perform the linear calculations using NIMROD as the first step. This work contributes to the TAE benchmark activity proposed by the ITPA Energetic Particle Topical Group\cite{Mishchenko2009}. Both the excitation of TAEs and their transition to energetic particle modes (EPMs) have been obtained. The dependence of frequency and growth rate of TAEs on temperature and density of energetic particles are consistent with the calculations from the eigenvalue code NOVA-K\cite{Konies2012}, as well as the gyrokinetic code GYGLES\cite{Mishchenko2009}. We also find that it is the relative EP pressure gradient at the radial location of TAE gap, which represents the drive strength of EPs, that strongly affects the growth rate of TAEs. From the calcultions, the conditions for the transition of TAE to EPM are further identified.

The rest of paper is organized as follows. In Section II, we describe the HK-MHD model used in NIMROD calculations. We introduce the setup of calculations, as well as the Alfv\'en continuum and the energetic particles density profiles in Section III. In Section IV, the different effects of EPs on TAEs, including the temperature, density, density gradient and position of maximum drive strength effects, are studied. The summary and discussion are given in the final section.

\section{The simulation model}
The hybrid kinetic-MHD model implemented in the NIMROD code is used in our simulations. The background plasma and energetic ions are modeled using MHD equations and drift kinetic equations, respectively\cite{Kim2004,Kim2008}. The resistive two-fluid MHD equations are solved as an initial-boundary value problem that is decretized on a mesh of finite elements in the poloidal plane and with a finite Fourier series in the toroidal direction\cite{Sovinec2003,Sovinec2004}. In this work, we use the ideal MHD equations
\begin{align}
\frac{\partial \rho}{\partial t} + \nabla\cdot(\rho\bm V) &= 0 ,\\
\rho\left(\frac{\partial \bm V}{\partial t}+\bm V\cdot\nabla\bm V\right) &=
\bm J\times\bm B-\nabla p_b-\nabla\cdot\bm P_f \label{eq:momen} ,\\
\frac{1}{\gamma_h-1}\left(\frac{\partial p}{\partial t} +
  \bm V\cdot\nabla p\right) &=
-p\nabla\cdot\bm V ,\\
\frac{\partial \bm B}{\partial t} &= -\nabla\times \bm E ,\\
\bm J &= \frac{1}{\mu_0}\nabla\times\bm B ,\\
\bm E + \bm V \times\bm B &= 0 ,
\end{align}
where subscripts $b,f$ denote bulk plasma and fast particles, $\rho,\bm V$ is fluid element density and velocity for the bulk plasma, neglecting the contribution of fast particles, $p$ is the pressure of entire plasma, $p_b$ is the pressure of bulk plasma, $\bm P_f$ is the pressure tensor of fast particles, and $\gamma_h$ is ratio of specific heats. Other notations are all conventional.

In HK-MHD model, it is assumed that the density of fast species is much lower than that of bulk plasmas but the fast species pressure is on the order of the bulk plasma pressure, i.e. $n_f \ll n_b$
and $\beta_f\sim \beta_b$, and $\beta\equiv 2\mu_0p/B^2$
is the ratio of thermal energy to magnetic energy. In this approximation,
we neglect the contribution of energetic particles to the center of mass velocity.
If we take the center of the mass velocity of energetic ions to be zero,  $\bm P_f$ in the momentum equation [Eq. \eqref{eq:momen}] can be calculated from the distribution function and the velocity of energetic ions,
\begin{equation}
  \bm P_f = m_f \int \bm v_f \bm v_f f_f(\bm r_f,\bm v_f) d^3 v_f \label{eq:fmom} ,
\end{equation}
where $m_f$, $\bm r_f$ and $\bm v_f$ are the mass, the spatial coordinate vector and the velocity of fast ions.\\

The $\delta f$ PIC method is used to solve the drift kinetic equation of energetic particles.
In the limit of strong magnetic field, the drift kinetic approximation
reduces the 6D phase space $(\bm r,\bm v)$ to 4D $(\bm r, v_{\parallel},\mu)$
with one adiabatic invariant (i.e. the first adiabatic invariant
$\mu = \frac 12 mv_{\bot}^2/B$).
If we substitute $f_f=f_{f0}+\delta f_f$
into Eq. \eqref{eq:fmom}, where $f_{f0}$ and $\delta f_f$ are the equilibrium
and the perturbed distribution function of fast particles, respectively, then $\bm P_f$ can be calculated as following
\begin{equation}
  \bm P_f =\bm P_{f0} + \delta\bm P_f ,
\end{equation}
\begin{equation}
  \delta\bm P_f=m_f \int \bm v_f \bm v_f \delta f_f(\bm r_f,\bm v_f) d^3 v_f ,
\end{equation}
where $\bm P_{f0}$ and $\delta\bm P_f$ are the equilibrium and the perturbed fast particle pressure tensor, respectively.
The condition for the force balance in equilibrium is given by
\begin{equation}
  \bm J_0\times\bm B_0 = \nabla p_0 + \nabla p_{f0} ,
\end{equation}
where the assumption is that the anisotropic components of fast particle
pressure tensor in equilibrium are zero and the tensor $\bm P_{f0}$ is reduced to a scalar
$p_{f0}$. Note that the steady state fields satisfy a scalar pressure force
balance, which is based on the assumption that the form of equilibrium energetic particle distribution
is isotropic in velocity space. With the solution for $\delta f_f$, we can calculate the pressure tensor. In the drift-kinetic approximation, the CGL-like pressure tensor can be used, $\delta \bm P_f = \delta p_{\bot} \bm I + (\delta p_{\parallel}-\delta p_{\bot}) \bm b\bm b$, where $\delta p_{\bot} = \int \mu B\delta f_f d^3 v_f$, $\delta p_{\parallel} = \int v_{\parallel}^2\delta f_f d^3v_f$. In the implementation, $\bm I$ is the unit tensor, and $\bm b=\bm B/B$.

\section{The simulation setup}
\begin{figure}
\includegraphics[width=0.7\textwidth]{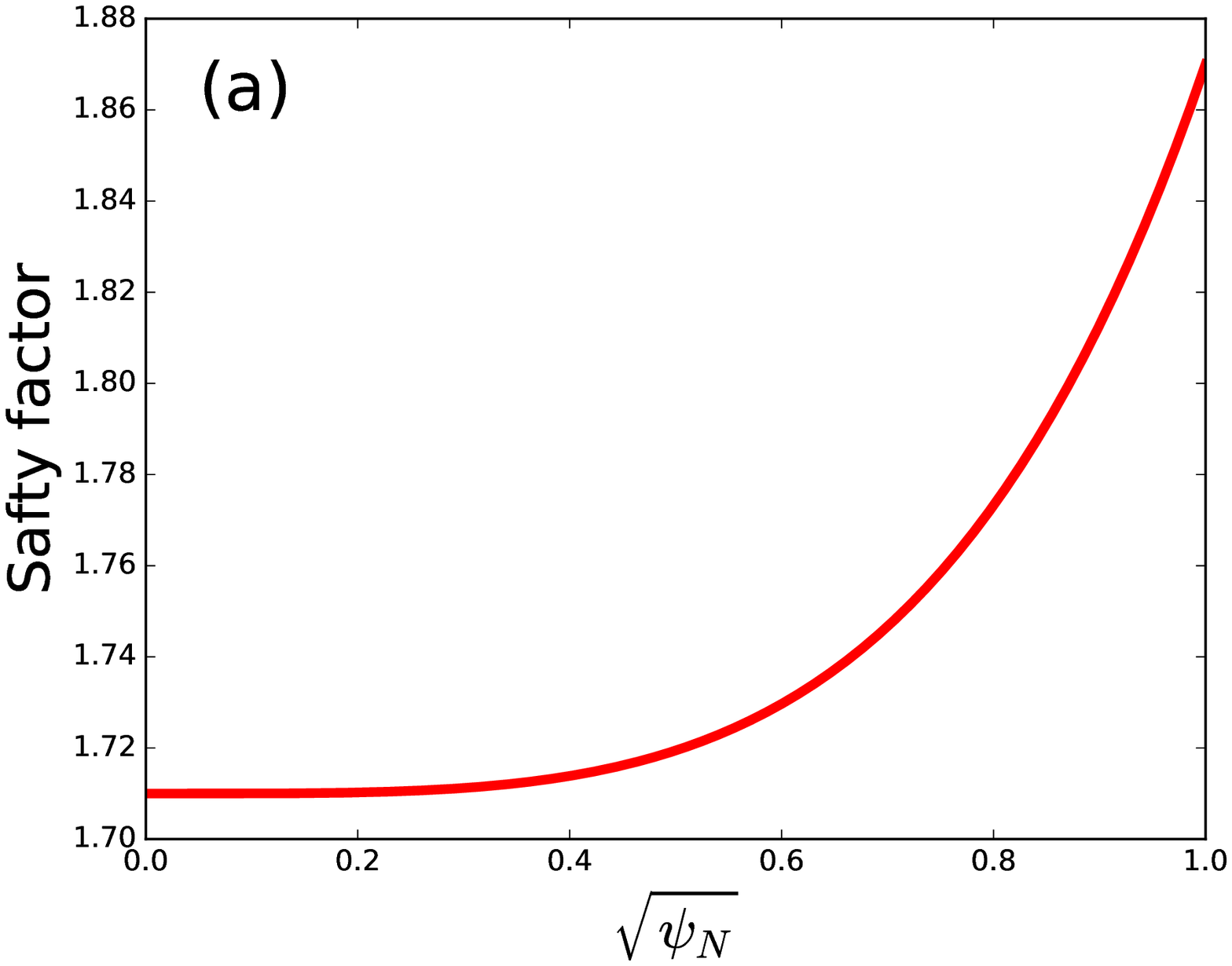}
\includegraphics[width=0.7\textwidth]{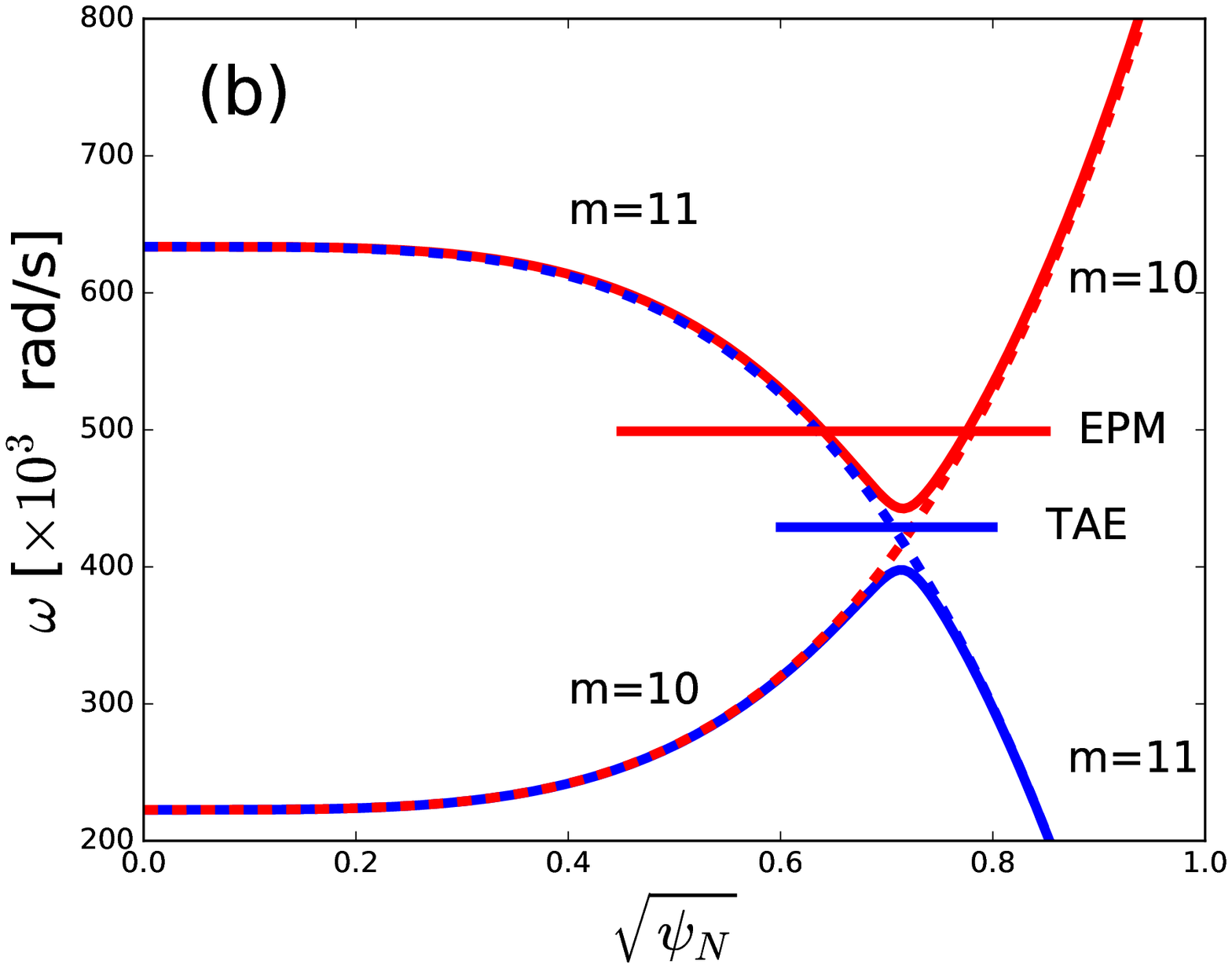}
\caption{Safety factor profile (a) and Alfv\'en wave continuum (b)
for toroidal mode number $n=6$ with poloidal mode numbers $m=10$ and $m=11$.
The dotted lines are from the cylindrical geomety. The 2D mode structures of TAE and EPM can be found in Fig.\ \ref{Fig3} (a) and (b). \label{Fig1}}
\end{figure}

In the ITPA benchmark case, circular cross section tokamak with the major radius $R_0=10$ m and the minor
 radius $a=1$ m is considered\cite{Konies2012, Mishchenko2009}. The radial profile of safety factor is
$q=1.71+0.16(r/a)^4$ (Fig.\ \ref{Fig1} (a)), where $r$ is the radial coordinate for flux surface.
The background number densities of hydrogen ions and electrons are assumed satisfy quasi-neutral condition.

The uniform radial profiles for both density and temperature of background plasma are used, where the ion density is $n_i=2\times 10^{19} m^{-3}$,
electron and ion temperatures are $T_e=T_i = 1\, keV$.
The background plasma $\beta=2\mu_0(n_iT_i+n_eT_e)/B^2\approx0.184\%$.
The normalized poloidal flux function $\sqrt{\psi_N}=\sqrt{\psi/\psi_a}$, where $\psi_a$ is the value of poloidal flux $\psi(r)$ at boundary $r=a$.
Energetic particle distribution in velocity space is Maxwellian
\begin{equation}
  f_M = 4\pi v_f^2\sqrt {\left(\frac{m_f}{2\pi T_f}\right)^3} e^{-\frac{m_fv_f^2}{2T_f}} ,
\end{equation}
where $m_f$, $v_f$ and $T_f$ are the mass, the velocity and the temperature of fast ions.

\subsection{Alfv\'en continuum}
As shown in Fig.\ \ref{Fig1} (b), the ideal MHD SAWs continuum can be obtained from the following dispersion relation\cite{Fu1989a}, which is valid for the low $\beta$ plasmas in the large aspect ratio limit,
\begin{equation}
  \omega_{\pm}^2=\frac{k_{\parallel m}^2v_A^2+k_{\parallel m+1}^2v_A^2\pm\sqrt{
      (k_{\parallel m}^2v_A^2-k_{\parallel m+1}^2v_A^2)^2+4\epsilon^2x^2k_{\parallel m}^2
  v_A^2k_{\parallel m+1}^2v_A^2}}{2(1-\epsilon^2x^2)} ,
\end{equation}
where $k_{\parallel m}=(n-m/q)/R_0$ is the parallel wavenumber, $R_0$ is the major
radius, $a$ is the minor radius, $q$ is the safety factor, $v_A=B/\sqrt{\mu_0\rho}$ is Alfv\'en velocity,
$\epsilon = 3a/2R_0$, $x=r/a$ is the normalized radius, $n$ is the toroidal mode number and $m$ is the poloidal mode number. For $n=6$, toroidicity induces coupling of $m = 10$ and $m = 11$ modes, and a TAE gap is formed in the SAW spectrum at $\psi_N = 0.5$ (or $\sqrt{\psi_N} = 0.707$).

\subsection{Energetic particle density profile}
\begin{figure}[p!]
\includegraphics[height=9cm,width=14cm]{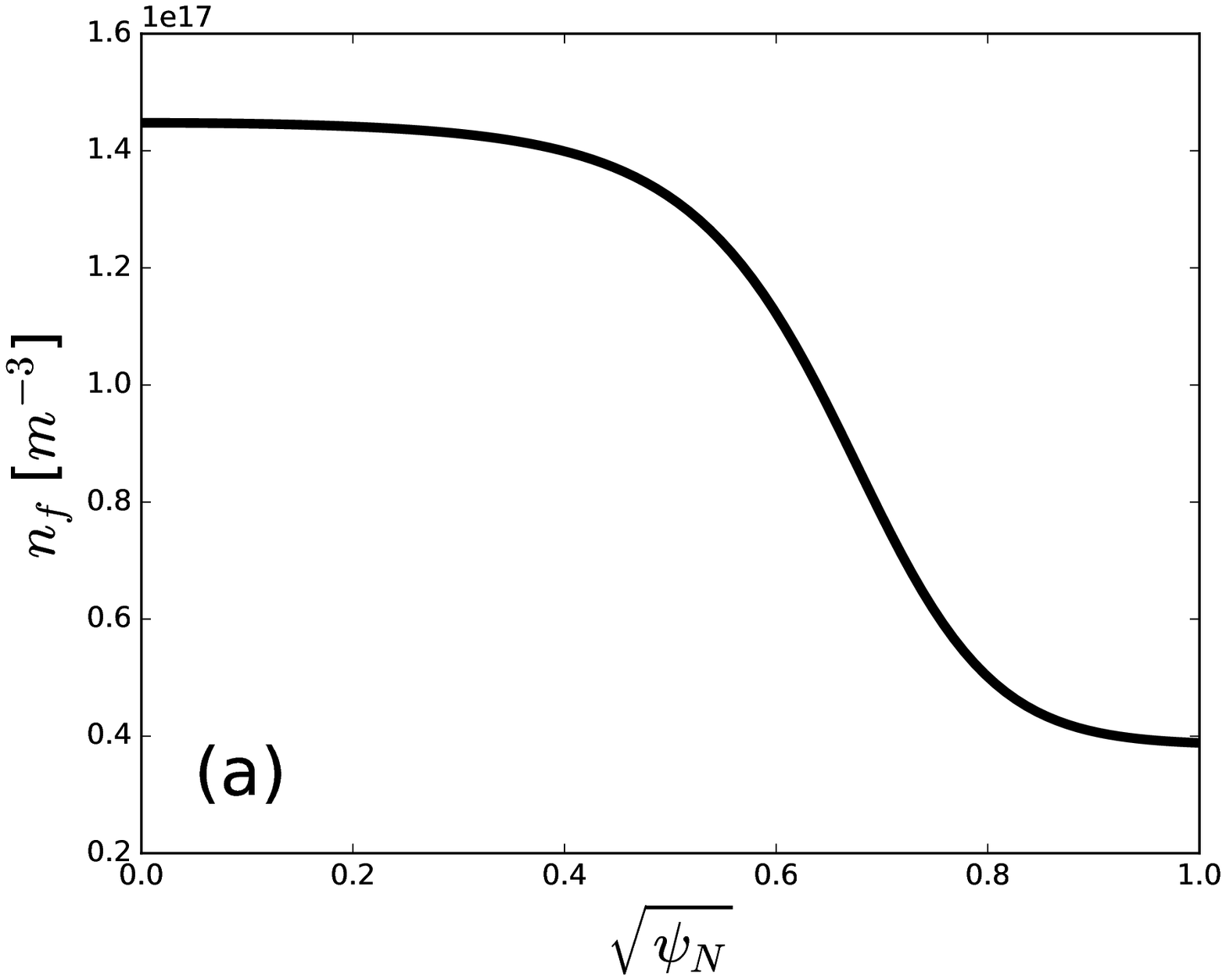}
\includegraphics[height=9cm,width=14cm]{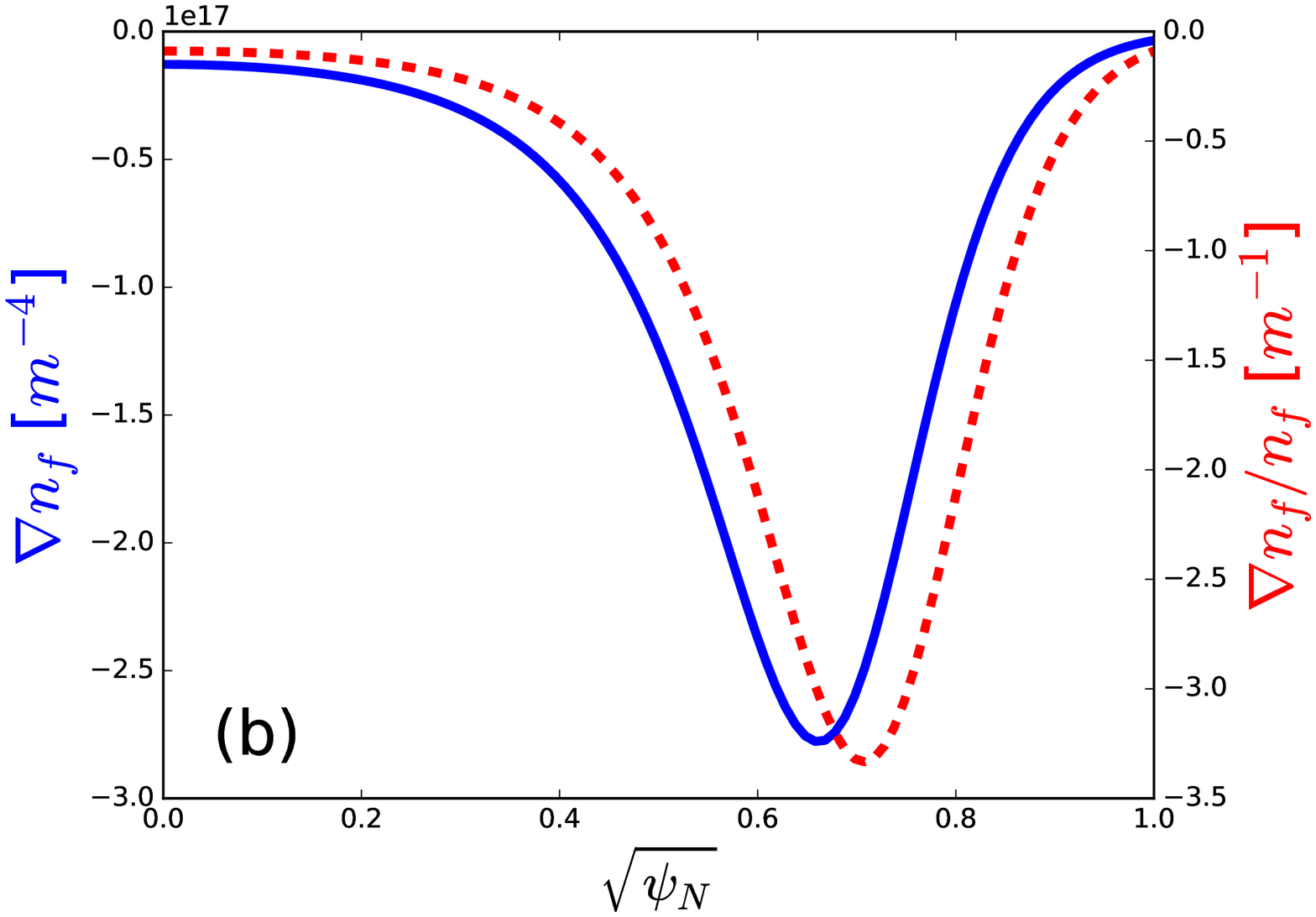}
\caption{Radial profiles of energetic particle density (a)
 and its radial gradient (b). The blue solid line represents the density gradient $\nabla n_f$ and the red dashed line represents the relative density gradient $\nabla n_f/n_f$. The unit of density is $m^{-3}$.
\label{Fig2}}
\end{figure}

The fast particle temperature $T_f$ is uniform in space and the fast particle density
$n_f$ is given by
\begin{equation}
  n_f(\psi_N)=n_{0f}\exp\left[-\frac{\Delta_{nf}}{L_{nf}}\tanh
    \left(\frac{\psi_N-\psi_{N0}}{\Delta_{nf}}\right)\right] ,
\label{eq:n0f}
\end{equation}
where $n_{0f}$ is the fast ion density at $\psi_N=\psi_{N0}$. The position of the maximal relative gradient $|\nabla n_f|/n_f$ is chosen to be $\psi_{N0}=0.5$, the profile width $\Delta_{nf}=0.2$ and the profile length $L_{nf}=0.3$. As shown in Fig.\ \ref{Fig2}, with density parameter $n_{0f}$=$7.5\times10^{16}m^{-3}$, the energetic particle density drops from core to  boundary. The maximum density gradient is located around the radial position $\psi_N$=0.44 (or $\sqrt{\psi_N}$=0.66) and the maximum relative gradient $|\nabla n_f|/n_f$ is located at the radial position $\psi_N$ =0.5 (or $\sqrt{\psi_N}$=0.707) where the TAE gap is formed (Fig.\ \ref{Fig1} (b)).

\section{NIMROD simulation results}
\subsection{Effects of energetic particle temperature }
The contour plot of radial magnetic field $B_\psi$ in Fig.\ \ref{Fig3} (a) shows a typical TAE pattern for the base case with energetic particle density $n_{0f}$=$7.5\times10^{16}m^{-3}$ and temperature $T_f$=400 keV.
One can see that this structure corresponds to the coupling of $m=10$ and $m=11$ mode.
The frequency ($\omega$$\approx$$4.29\times10^5 rad/s$) and radial location ($\sqrt{\psi_N}$$\approx$$0.7$)
are located within the continuum gap predicted by the theory, as indicated in Fig.\ \ref{Fig1} (b).

We vary the temperature of energetic particle from 100 keV to 800 keV with fixed density ($7.5\times10^{16}m^{-3}$),
and calculate the frequency and growth rate of the excited unstable TAE (Fig.\ \ref{Fig4}).
The frequency is found to increases with temperature, whereas the growth rate increases first and then decreases.
The frequencies are located in the TAE gap for all cases (Fig.\ \ref{Fig4} (a)), suggesting the TAE nature of the excited modes. From the GYGLES simulation results including the FLR (Finite-Larmor-Radius) effects, the TAE frequency first decreases and then increases with temperature, which is slightly different from our finding when the temperature of EPs is very low. When the EP temperature is higher, the variations of the frequency from two codes are similar to each other. The growth rate from NIMROD simulations are consistent with the results from NOVA-K\cite{Konies2012} and GYGLES\cite{Mishchenko2009} (Fig.\ \ref{Fig4} (b)). The dependence of growth rate on EP temperature can be understood from the interaction between energetic particles and TAE. There is a resonant velocity of energetic particle at which the strongest interaction occurs between particles and wave because of the match between the EP and the wave phase velocities. For a fixed density of energetic particles, when the thermal velocity of EPs approaches the resonant velocity from below, the interaction becomes stronger, thus the growth rate increases. This corresponds to the initial increase of temperature of energetic particles, which is equivalent to the increase in EP pressure because of the fixed EP density. When the EP temperature hence EP pressure further increases, the thermal velocity of EPs surpasses the resonant velocity and resonant interaction becomes weaker, which leads to the decrease in growth rate. For this reason, the growth rate first increases and then decreases with the EP pressure, as shown in simulation results (Fig.\ \ref{Fig4} (b)).


We further investigate the effect of EP temperature with the EP pressure hence $\beta_f$ fixed. For the base case, the EP density $n_{0f}=7.5\times10^{16}m^{-3}$ and temperature $T_{f}=400keV$, $\beta_f=0.134\%$, and the background plasma $\beta_0=0.18\%$. Keeping $\beta_f$ constant, it can be found that in Fig.\ \ref{Fig5}, with the increase in EP temperature, the frequency increases first and then decreases. When the temperature is relatively small and the density is relatively big, the mode frequency is below the gap and it is EPM. As the temperature increases and the density decreases, the frequency increases to be located in the gap and the mode change from EPM to TAE. That is, there is a threshold ($1.5\times10^{17}m^{-3}$) for EP density, above which EPM can be excited even though the EP temperature is relativly small. Except when the EP temperature is close to zero, the variations of frequencies from NIMROD and GYGLES simulations are similar. The growth rate increases first and then decreases, and around $T_{f}=200keV$ it reaches to the maximum. The dispersion relation in tokamak with inhomogeneous magnetic field can be approximated as $\omega(r)=k_{\parallel}(r)V_A(r)$ which shows shear Alfv\'en waves with a radial extent are dispersive due to different phase velocities. Considering the magnetic curvature, the dispersion relation can be revised to $\omega(r)-k_{\parallel}(r)v_{\parallel}(r)-k_{\bot}(r)V_{di}(r)=0$, where $V_{di}$ is the ion magnetic drift velocity \cite{Rosenbluth1975}. The drift velocity term reduces effectively the eigenmode frequency so that the EP particles are able to resonate at the speed $v = V_A/3$.
As the Alfv\'en velocity $V_A$ ($B/\sqrt{\mu_0 \rho}$) at the radial location of the gap ($\approx 10.7m$) is $1.369\times10^7 m/s$, the corresponding resonant temperature of energetic particle is $217keV$ which is consistent with the location of maximum growth rate.

\subsection{Effects of energetic particle density}
As shown in Fig.\ \ref{Fig6}, for the fixed temperature of energetic particle, both the frequency and the growth rate increase with density. Comparing the simulation results from the NIMROD code with ZLR effect and the GYGLES code with FLR effects, one finds that the FLR effect only weakly enhances the frequency and suppresses the growth rate of EP-driven TAE.
When the density is relatively small, the frequency is located in the TAE gap,
but when the density increases to a certain value ($1.0\times10^{17}m^{-3}$), the frequency becomes higher above the upper value of gap so that TAE will transform into EPM. For the energetic particle density $n_{0f}=1.0\times10^{16}m^{-3}$ and the temperature $T_{f}=200keV$, the $\beta$ ratio of energetic particles to bulk plasma $\beta_f/\beta_0$ is 0.1. As the EP density increases to $1.0\times10^{17}m^{-3}$, $\beta_f$ approaches $\beta_0$. This is the threshold value
of energetic particle density above which the TAE-EPM transition occurs. If the energetic particle density continues to increase, the mode frequency increases further beyond the gap to an even larger value. That suggests that EPM can be excited when the energetic particle $\beta_f$ becomes larger than the bulk plasma $\beta$.



\begin{figure}[p!]
\includegraphics[height=9cm,width=12cm]{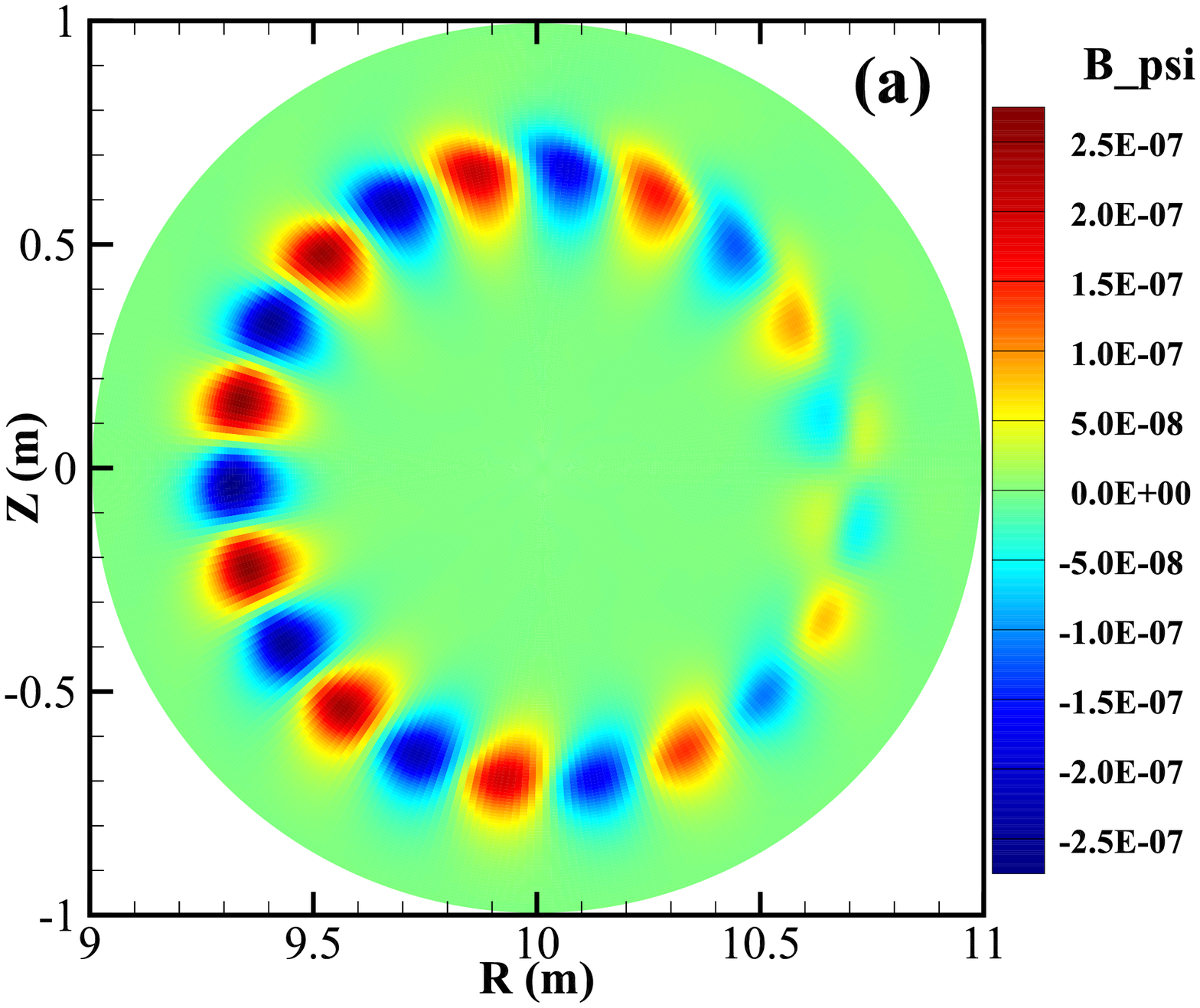}
\includegraphics[height=9cm,width=12cm]{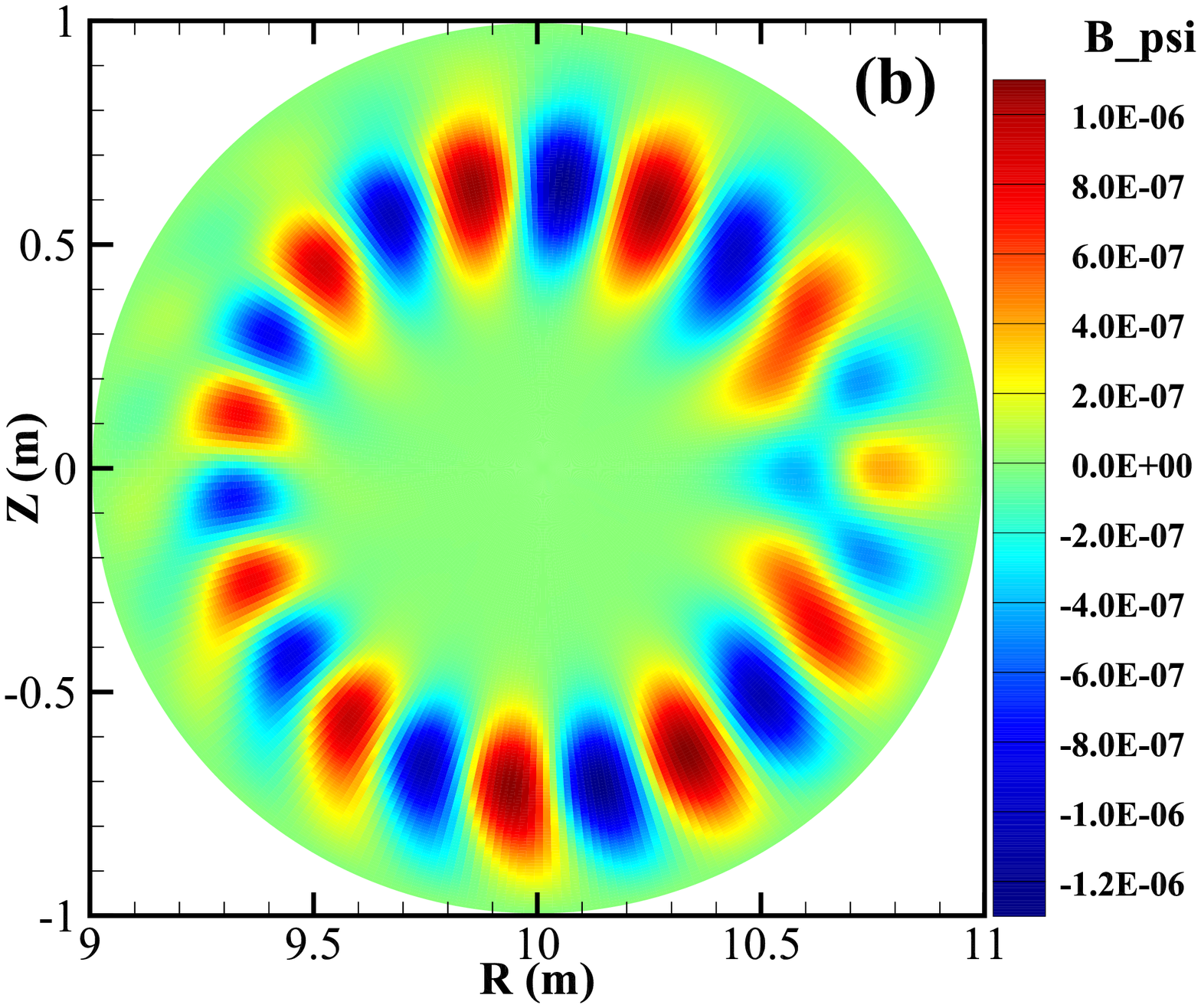}
\caption{Contours of the radial component of magnetic field $B_{\psi}$ for TAE (a) and EPM (b) with toroidal mode number $n=6$ and poloidal mode number $m=10$ and $11$. The locations of TAE and EPM in continuum can be found in Fig.\ \ref{Fig1} (b). The EP parameters are: temperature $T_f$=400, and (a) density $n_{0f}$=$7.5\times10^{16}m^{-3}$ in the TAE; (b) density $n_{0f}$=$8.0\times10^{17}m^{-3}$ in the EPM case. \label{Fig3}}
\end{figure}

\begin{figure}[p!]
\includegraphics[height=8cm,width=12cm]{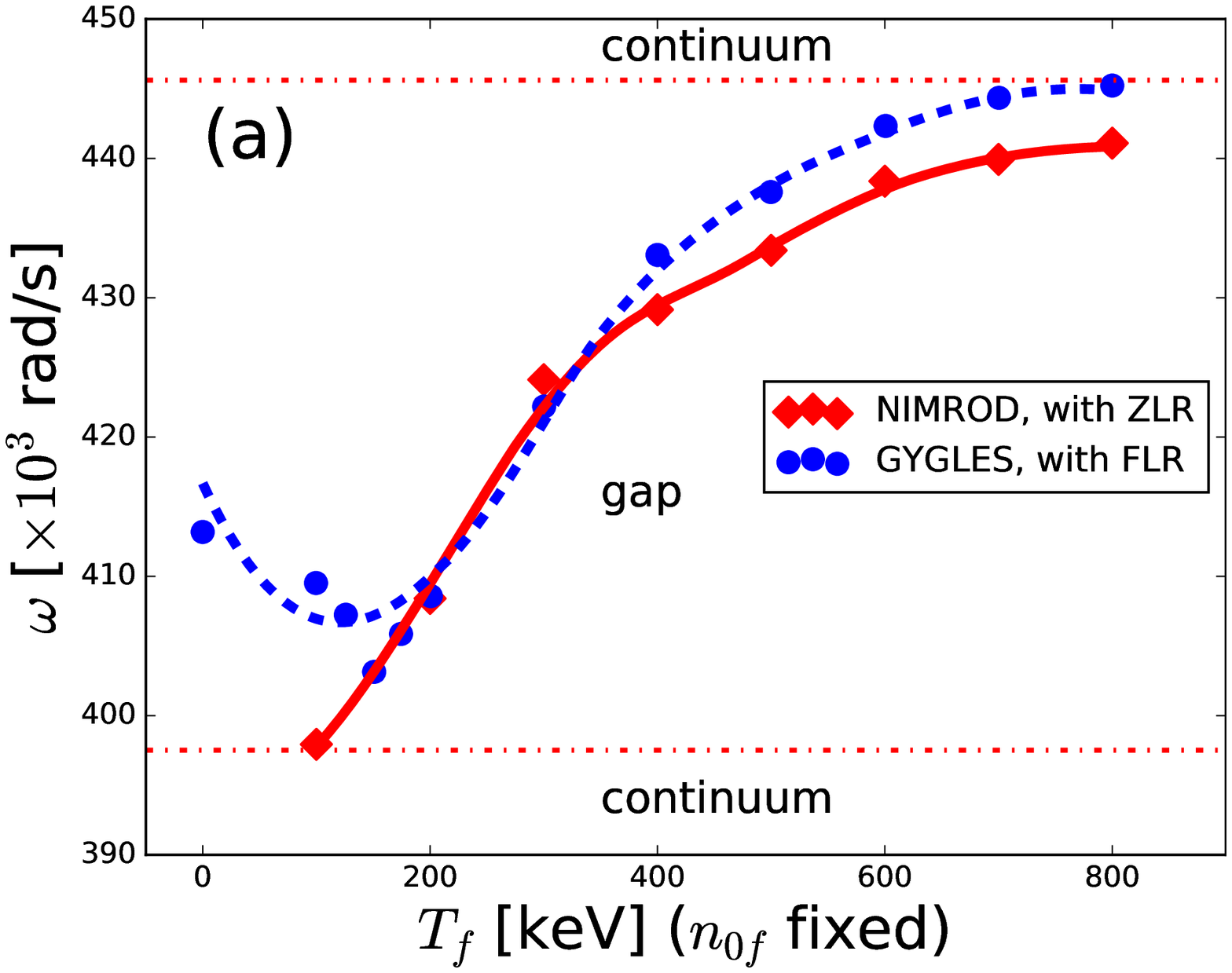}
\includegraphics[height=8cm,width=12cm]{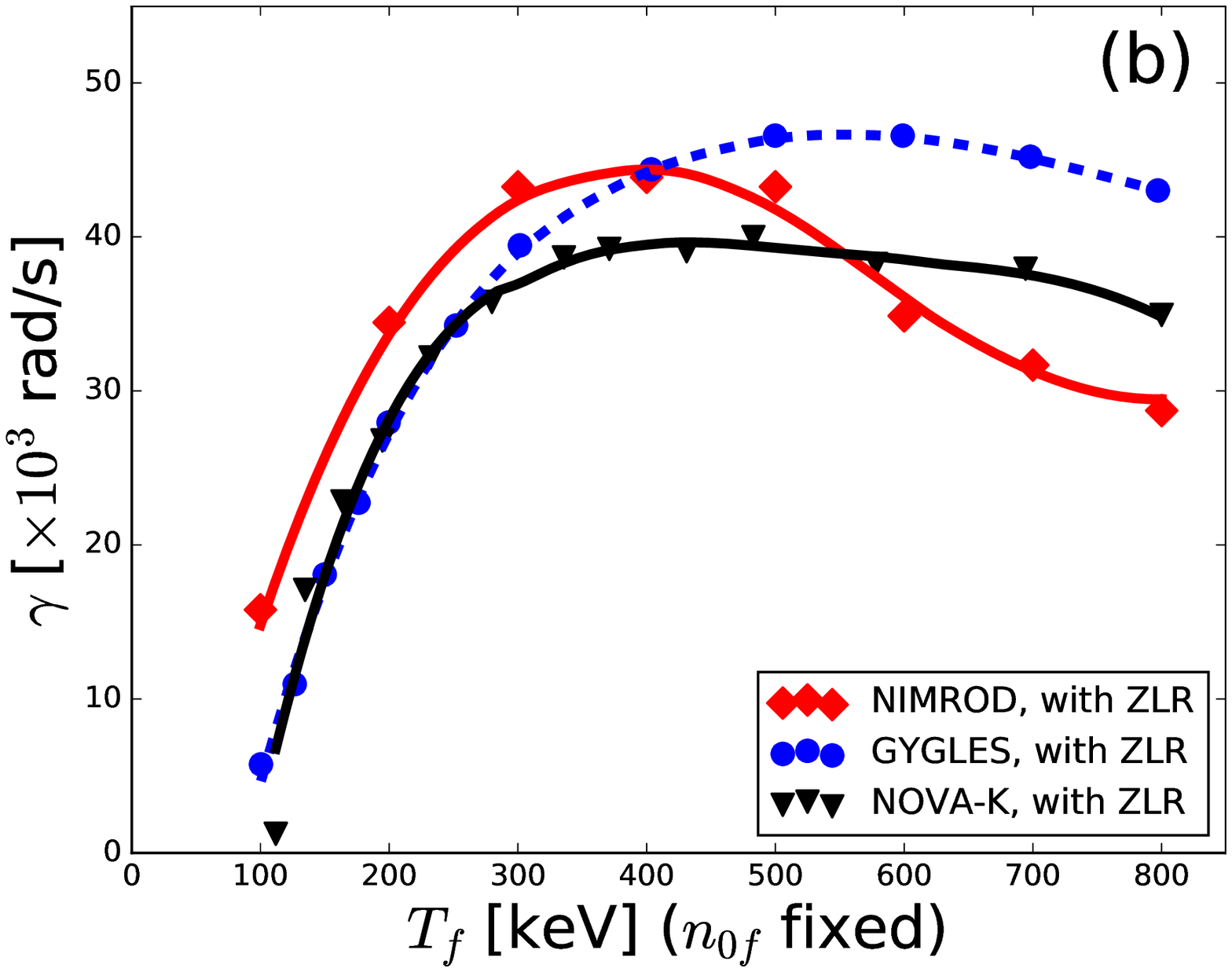}
\caption{The frequency (a)
and growth rate (b) of TAE mode as functions of the energetic particle temperature $T_f$.
ZLR stands for Zero Larmour Radius and FLR for Finite Larmour Radius. \label{Fig4}}
\end{figure}

\begin{figure}[p!]
\includegraphics[height=8cm,width=12cm]{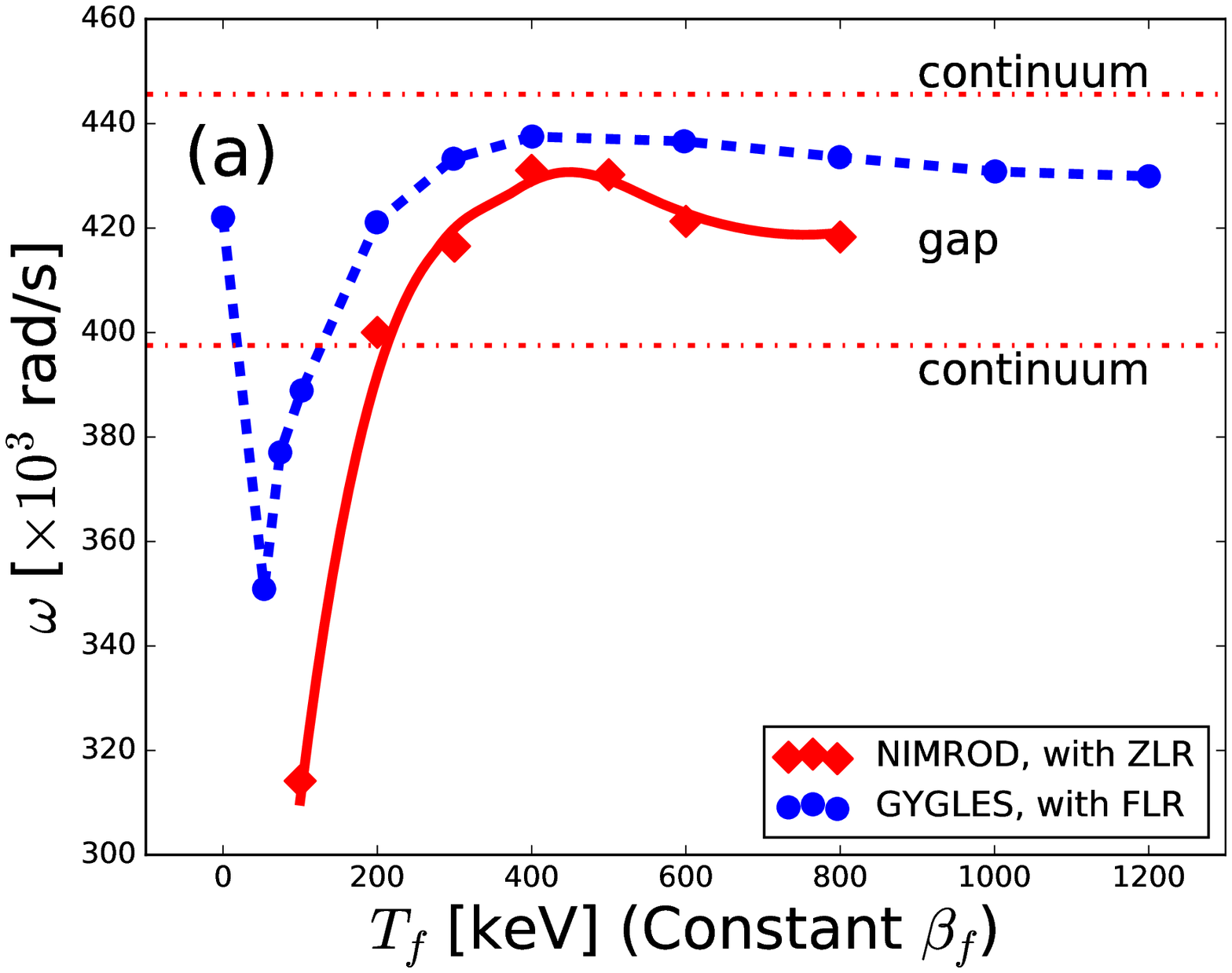}
\includegraphics[height=8cm,width=12cm]{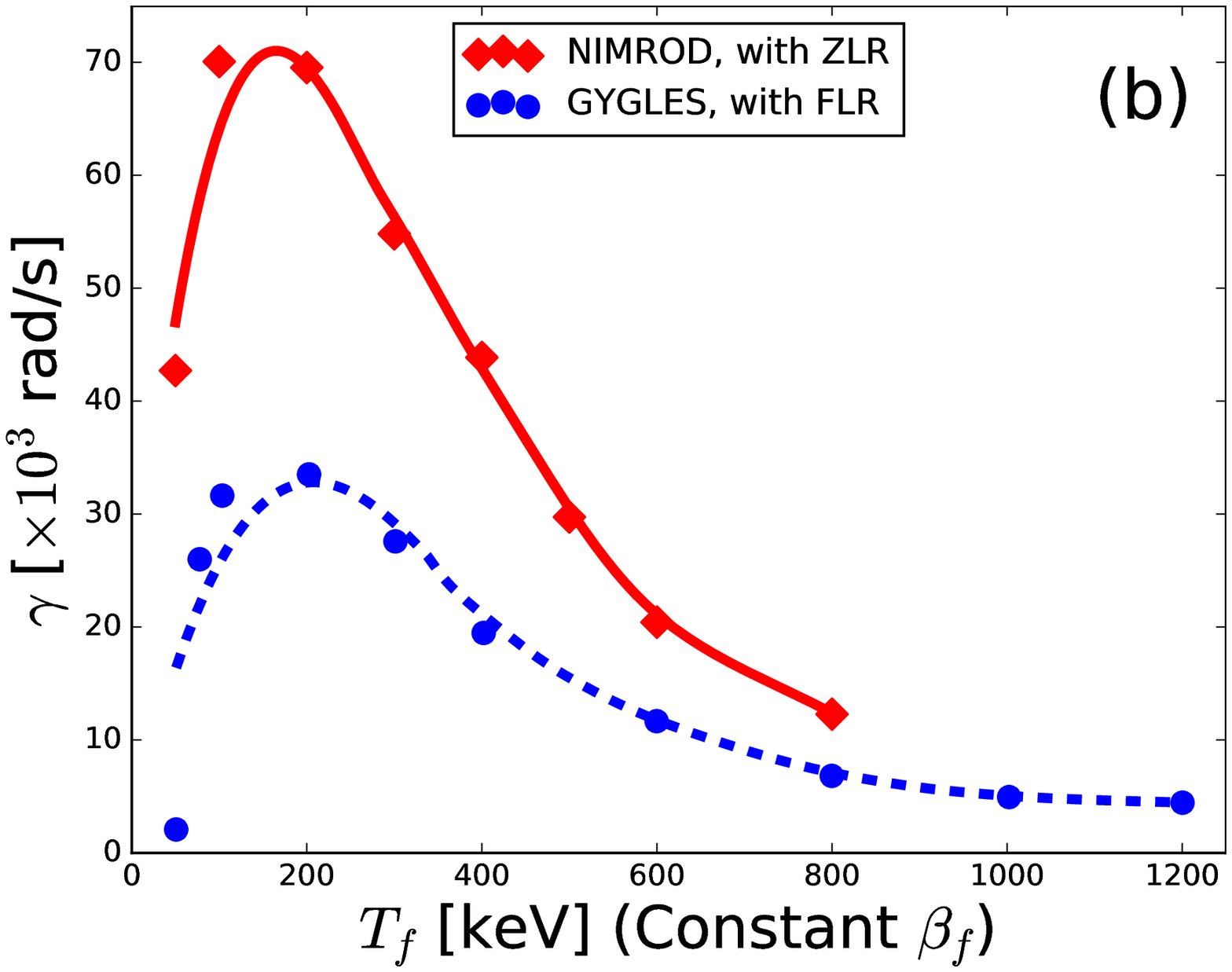}
\caption{The frequency (a)
and growth rate (b) of TAE mode as functions of the energetic particles temperature $T_f$.
The constant $\beta_f=0.134\%$. \label{Fig5}}
\end{figure}

\begin{figure}[p!]
\includegraphics[height=8cm,width=12cm]{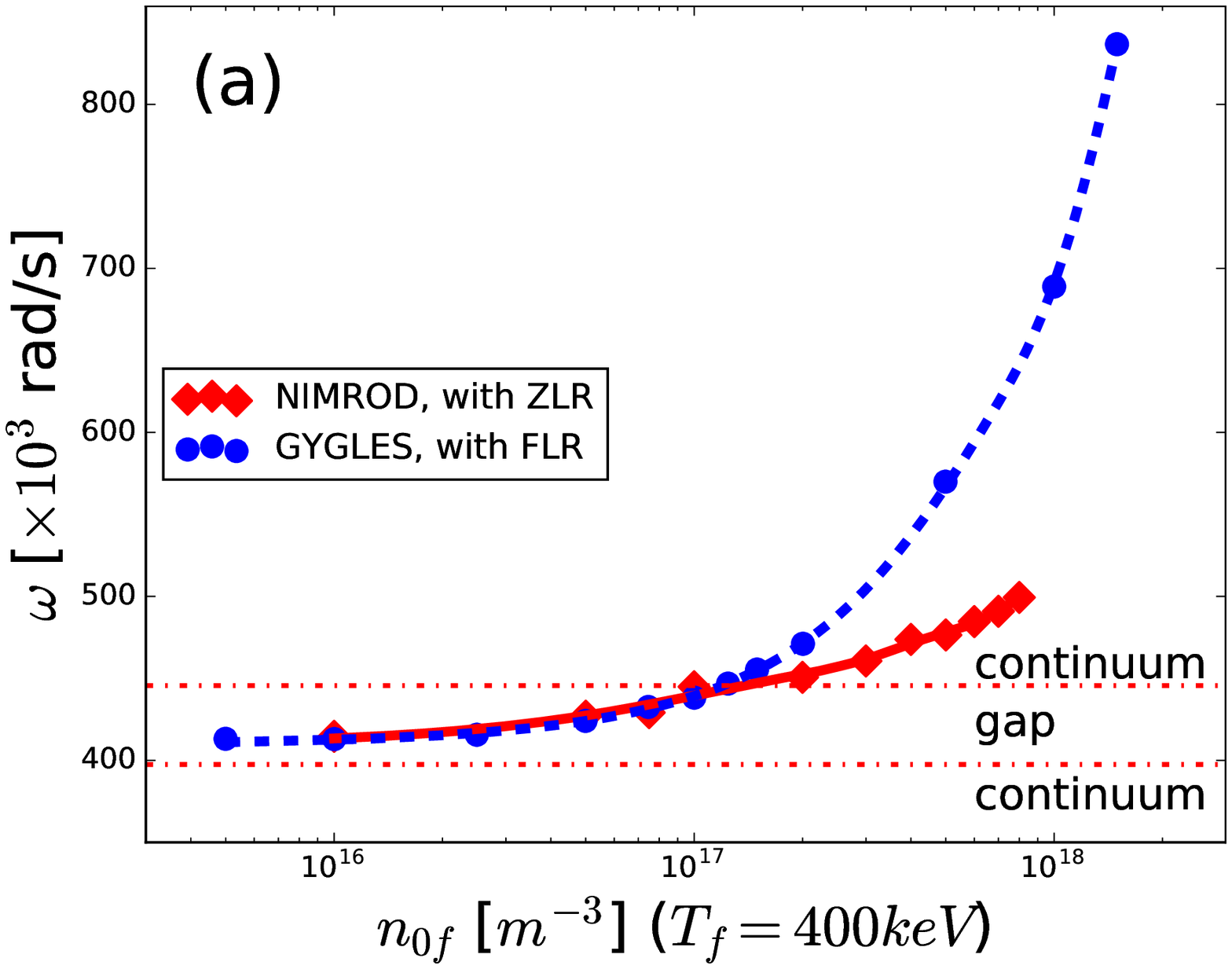}
\includegraphics[height=8cm,width=12cm]{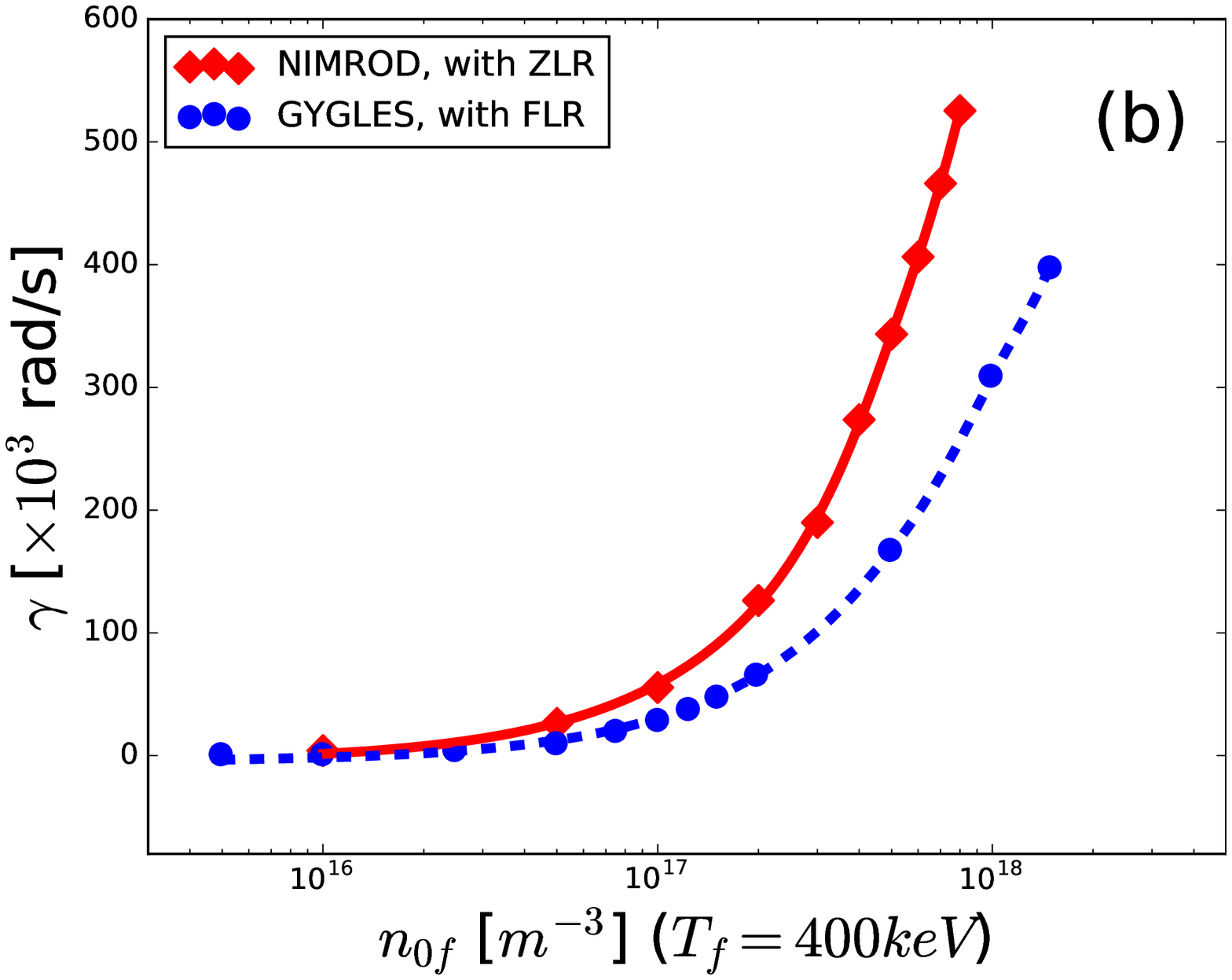}
\caption{The frequency (a)
and growth rate (b) of TAE mode as functions of the energetic particle density parameter $n_{0f}$.
The energetic particle temperature $T_f$ is fixed to be $400keV$. \label{Fig6}}
\end{figure}

The TAE-EPM transition can be also observed from the changes in the dominant mode structure in poloidal plane (Fig.~\ref{Fig3}). When the EP density is $7.5\times10^{16}m^{-3}$, the TAE is located at a relatively narrow region around $\sqrt{\psi_N}$=0.707 between $\sqrt{\psi_N}$=0.6 and $\sqrt{\psi_N}$=0.8 (Fig.\ \ref{Fig3} (a)). When the EP density increases to $8.0\times10^{17}m^{-3}$, the EPM mode structure expands to a wider region around $\sqrt{\psi_N}$=0.707 from $\sqrt{\psi_N}$=0.45 to $\sqrt{\psi_N}$=0.85 (Fig.\ \ref{Fig3} (b)).


\subsection{Effects of energetic particle density gradient}
\begin{figure}[p!]
\includegraphics[height=8cm,width=12cm]{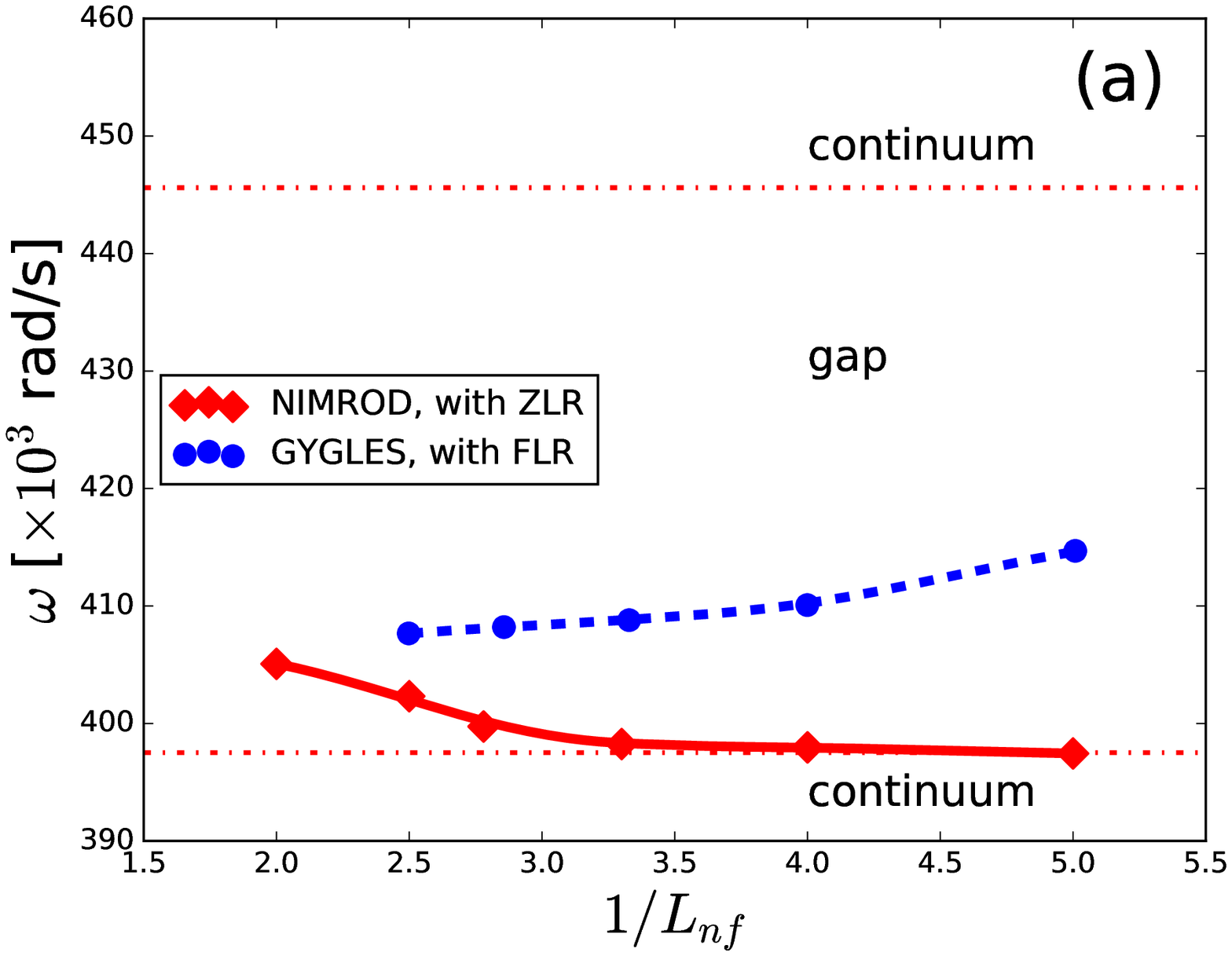}
\includegraphics[height=8cm,width=12cm]{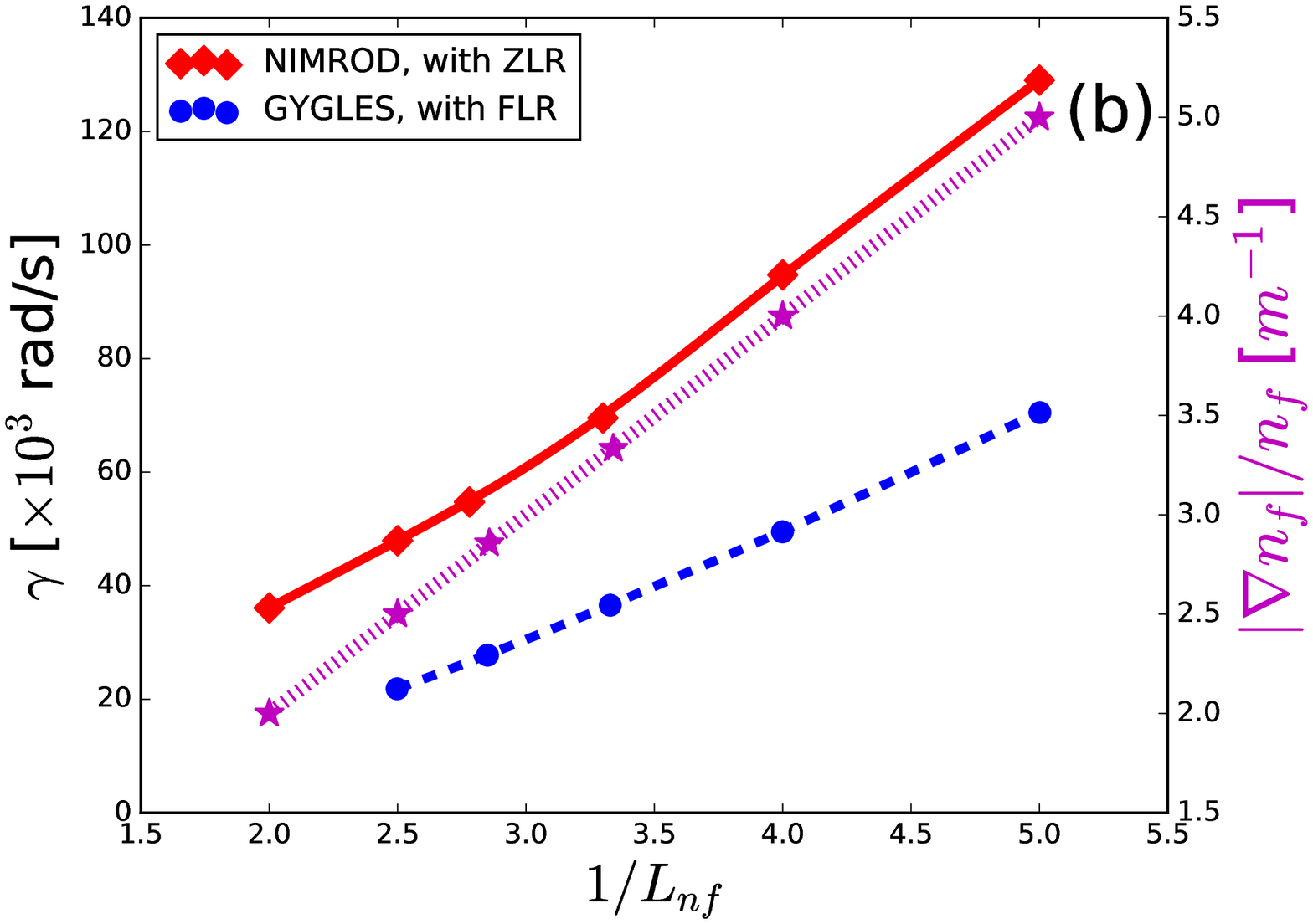}
\caption{The frequency (a)
and growth rate (b) of TAE mode as functions of the energetic particle density gradient parameter $L_{nf}^{-1}$.
The magenta dotted line represents the absolute value of relative density gradient $|\nabla n_f|/n_f$ at the location of TAE gap $\psi_N=0.5$.
The energetic particle density parameter $n_{0f}$=$1.5\times10^{17}m^{-3}$ and temperature $T_f$=200 KeV are fixed. \label{Fig7}}
\end{figure}

In order to investigate the effect of energetic particle density gradient, we vary the parameter $L_{nf}$ used to specify the EP density profile in Eq.~(\ref{eq:n0f}). In the base case, EP density parameter $n_{0f}$=$1.5\times10^{17}m^{-3}$ and temperature $T_f$=200 KeV are selected where the growth rate is relatively large (Fig.\ \ref{Fig5}). Simulations indicate that the growth rate is nearly proportional to the absolute value of relative density gradient $|\nabla n_f|/n_f$ at the location of TAE gap $\psi_N=0.5$ (Fig.\ \ref{Fig7} (b)). The frequency first decreases and then increases with density gradient and it remains located within the TAE gap for most cases (Fig.~\ref{Fig7} (a)). Meanwhile, the growth rate increases monotonically. In contrast, the increase of energetic particle density gradient does not affect the dominant mode feature, even as the mode nature transits from TAE towards EPM.

\subsection{Radial location of maximum EP drive}
\begin{figure}[p!]
\includegraphics[height=8cm,width=12cm]{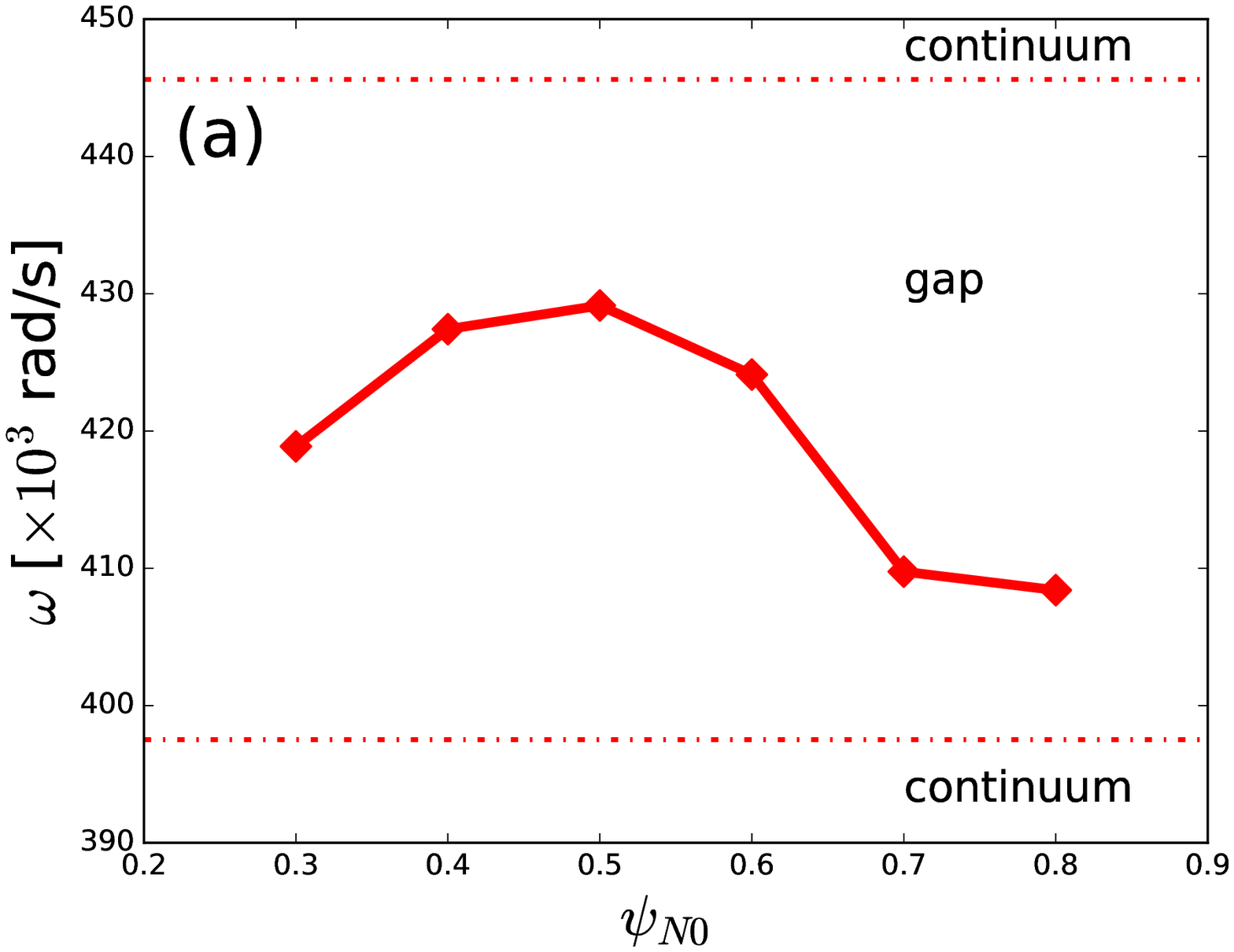}
\includegraphics[height=8cm,width=12cm]{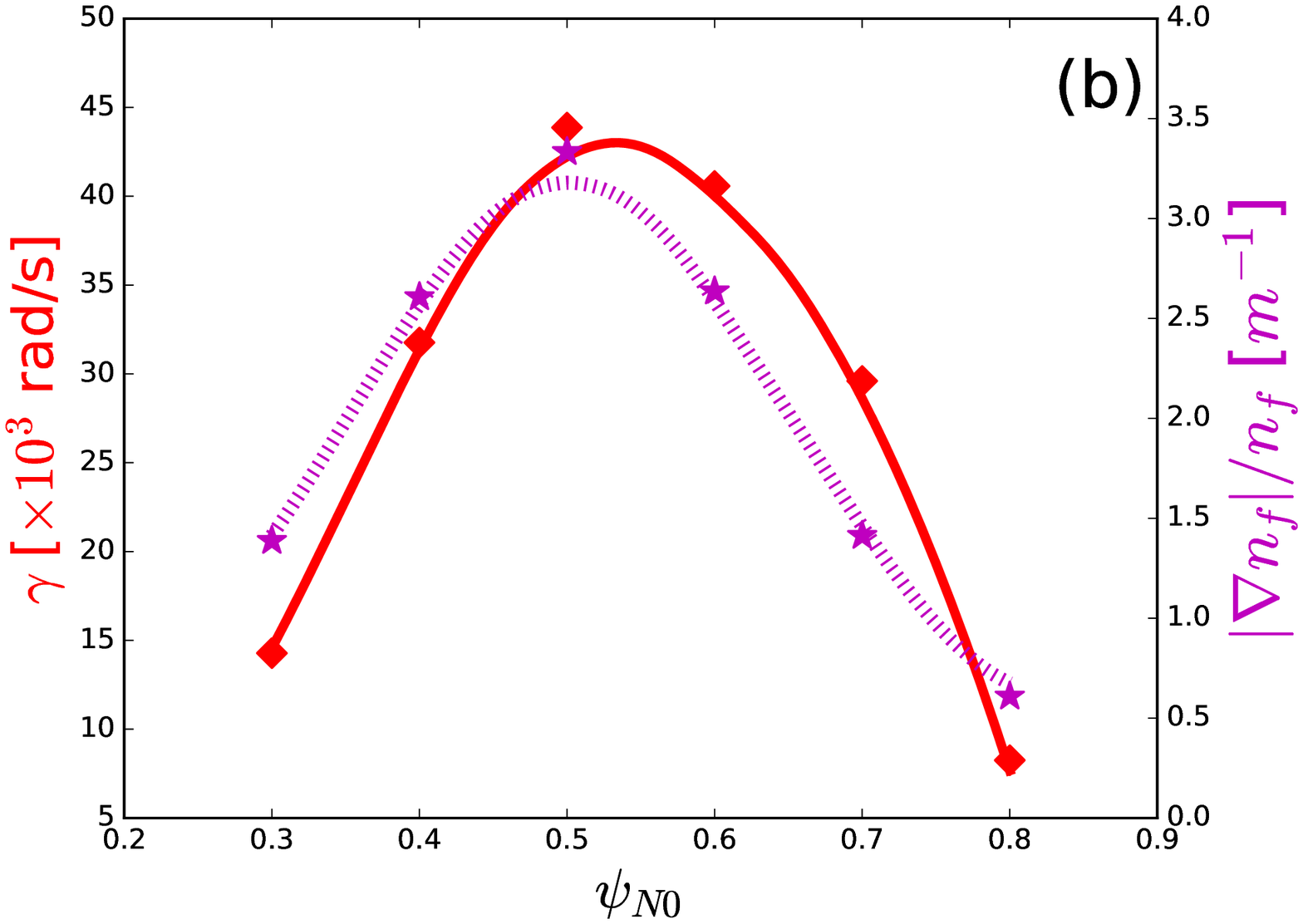}
\caption{The frequency (a)
and growth rate (b) of TAE mode as functions of the location of the maximum relative density gradient $\psi_{N0}$.
The magenta dotted line represents the absolute value of relative density gradient $|\nabla n_f|/n_f$ at the location of TAE gap $\psi_N=0.5$.
The energetic particle density parameter $n_{0f}$=$7.5\times10^{16}m^{-3}$ and temperature $T_f$=400 KeV are fixed. \label{Fig8}}
\end{figure}

The strength of energetic particle drive depends on the radial location of the maximum relative pressure gradient\cite{Wang2013}. In order to investigate the effect of energetic particle density gradient, we vary the parameter $\psi_{N0}$ in the EP density profile function in Eq.~(\ref{eq:n0f}), which represents the location of the maximum relative pressure gradient. EP density parameter $n_{0f}$=$7.5\times10^{16}m^{-3}$ and temperature $T_f$=400 KeV are fixed. Both the frequency and the growth rate first increase and then decrease with $\psi_{N0}$, and they reach the maximums at $\psi_{N0}=0.5$ (Fig.\ \ref{Fig8}). One can see that the absolute value of relative density gradient $|\nabla n_f|/n_f$ at the location of TAE gap $\psi_N=0.5$ has a positive correlation to the growth rate
and reaches the maximum value at $\psi_{N0}=0.5$ too (Fig.\ \ref{Fig8} (b)). For the case $\psi_{N0}=0.5$, the maximum relative density gradient collocates with the location where the $n=10$ and $n=11$ modes interact to form the TAE gap. That is, for the uniform temperature, when the radial location of the maximum relative density and the TAE gap overlap with each other, the EP drive becomes maximal. This result is consistent with the conclusion from previous GTC simulation studies on the DIII-D tokamak\cite{Wang2013}, that the TAE appears in the radial location where the relative pressure gradient of EPs is the strongest.

\section{Summary and discussion}
NIMROD simulations of the SAWs excitied by energetic particles in a circularly shaped tokamak
have been presented. The dependence of both the frequency and the growth rate on the energetic particle parameters
are studied, and the NIMROD results are consistent with those from NOVA-K \cite{Konies2012} and GYGLES \cite{Mishchenko2009}. When the $\beta$ value of energetic particles (i.e. $\beta_f$) becomes larger than that of bulk plasma (i.e. $\beta_0$), or when EP density exceeds a certain threshold with a relatively lower temperature, the TAEs can transit to EPMs. For the resonant thermal velocity ($v_A/3$) of energetic particles, the interaction between SAWs and EPs is the strongest. The density gradient at the radial location of TAE gap can strongly affect the growth rate of TAE, though not its frequency. The EP drive becomes the strongest when the maximum relative pressure gradient of EP is located at the radial position of TAE gap.

So far, the FLR effect is not considered in our simulation. In the future, the FLR and FOW (Finite-Orbit-Width) effects can be included and examined. In addition, for tokamak experiments, the strong interaction of EPs with background plasma may better be modeled with the slowing-down distribution function. To study EP driven TAEs and EPMs in realistic tokamak experiments, the slowing-down distribution function or the distributions obtained directly from experiments for EPs will be considered in our future simulation work.

\begin{acknowledgments}
This work was supported by National Magnetic Confinement Fusion Science Program of China under Grant Nos. 2014GB124002 and 2015GB101004, the National Natural Science Foundation of China grant No. 11205194, U.S. Department of Energy Grant Nos. DE-FG02-86ER53218 and DE-FC02-08ER54975, and the 100 Talent Program of the Chinese Academy of Sciences. This research used the computing resources from the Supercomputing Center of University of Science and Technology of China, and the National Energy Research Scientific Computing Center, a DOE Office of Science User Facility supported by the Office of Science of the U.S. Department of Energy under Contract No. 13DE-AC02-05CH11231.
\end{acknowledgments}

%



\newpage

\newpage

\end{document}